\documentclass{aa}
\usepackage{natbib}
\bibpunct{(}{)}{;}{a}{}{,}
\usepackage{graphicx}
\usepackage[figuresright]{rotating}
\usepackage{txfonts}
\begin{document}
\newcommand{\water}{H$_2$O}
\newcommand{\solum}{\hbox{L$_{\sun}$}}
\newcommand{\wlum}{$L_{{\rm H_2 O}}$}
\newcommand{\kms}{km s$^{-1}$}
\newcommand{\HII}{\hbox{H{\sc ii}}}
\newcommand{\HI}{\hbox{H{\sc i}}}

\defcitealias{newmas}{HPT}
    \title{New {\water} masers in Seyfert and FIR bright galaxies. II.
          }
   \subtitle{The intermediate luminosity range}

 \author{P.\ Castangia\inst{1,2}
         \and A.\ Tarchi\inst{2,3}
         \and C.\ Henkel\inst{1}
         \and K.\ M.\ Menten\inst{1}}

    \offprints{P.\ Castangia}

 \institute{Max-Planck-Insitut f\"ur Radioastronomie, Auf dem H\"ugel 69, 53121 Bonn, Germany\\
               \email{pcastang@mpifr-bonn.mpg.de}
               \and INAF-Osservatorio Astronomico di Cagliari, Loc. Poggio dei Pini, Strada 54, I-09012 Capoterra (CA), Italy
               \and INAF-Istituto di Radioastronomia, Via Gobetti 101, 40129 Bologna, Italy}
   
    \date{Received ; accepted }


  \abstract
  {Recently, a relationship between the water maser detection rate and far infrared (FIR) flux density has been found as a result of a 22\,GHz maser survey in a sample comprised of northern galaxies with 100\,$\mu$m flux density $>$\,50\,Jy and a declination $>-30${\degr}.}
  {The survey has been extended toward galaxies with lower FIR flux densities in order to confirm this correlation and to discover additional maser sources for relevant follow-up interferometric studies.}
   {A sample of 41 galaxies with 30\,Jy\,$< S_{\rm 100\,\mu m} <$\,50\,Jy and  $\delta > -30${\degr} was observed with the 100-m telescope at Effelsberg in a search for the 22\,GHz water vapor line. The average 3$\sigma$ noise level of the survey is 40 mJy for a 1\,{\kms} channel, corresponding to a detection threshold for the isotropic maser luminosity of $\sim$\,0.5\,{\solum} at a distance of 25 Mpc.}
   {Two detections are reported: a megamaser with an isotropic luminosity, {\wlum}, of $\approx$ 35\,{\solum} in the Seyfert/{\HII} galaxy NGC\,613 and a kilomaser with {\wlum}\,$\approx$\,1\,\solum\ in the merger system NGC\,520. The high luminosity and the presence of a Seyfert nucleus favor an association for NGC\,613 with an active galactic nucleus. The kilomaser in NGC\,520 was also detected with the Very Large Array, providing a position with subarcsecond accuracy.  The {\water} emission, originating from a $\la 0.02$\,pc sized region with a brightness temperature $\ga 10^{10}$\,K (if the observed variations are intrinsic to the masing cloud(s)), is close to one of the two radio continuum sources located in the inner parsecs of NGC\,520. The maser is most likely associated with a young supernova remnant (SNR), although an association with a low-luminosity AGN (LLAGN) cannot be ruled out. The maser detection rate, with 2 new maser sources out of 41 galaxies observed, is consistent with expectations extrapolated from the statistical properties of the $S_{\rm 100\,\mu m} >$\,50\,Jy sample. The {\water} kilomasers are ``subluminous'', while {\water} megamasers tend to be ``superluminous'' with respect to the FIR luminosity of their parent galaxy, when compared with sites of massive star formation in the Milky Way.}
   {}


   \maketitle
%

\section{Introduction}

Water maser emission at 22\,GHz ($\lambda$\,$\sim$\,1.3\,cm) originates from the transition between the 6$_{16}$ and 5$_{23}$ rotational levels of ortho-{\water} and forms in dense ($n{\rm (H_2)}$\,$\ga$\,10$^{7}$\,cm$^{-3}$) and warm ($T_{\rm kin}$\,$\ga$\,400\,K) molecular gas \citep{hen_review}. The most plausible inversion mechanism is collisional pumping with H$_2$ molecules \citep{neufeld}. In the Galaxy, water maser sources are common in regions of star formation and in the circumstellar envelopes of mass-losing oxygen-rich evolved stars. Since 1977, when the first extragalactic maser was detected in the nearby spiral galaxy M33 \citep{m33}, this maser line has been observed in $\sim$\,100 external galaxies (Braatz 2007, priv. com.). Isotropic 22 GHz line luminosities of extragalactic {\water} masers, {\wlum}, range from 10$^{-2}$\,{\solum} (e.\,g., M33, \citealt{m33}; IC\,10, \citealt{hen86}; IC\,342, \citealt{ic342}) to 23,000\,{\solum} (SDSS\,J0804+3607, \citealt{barvai}); for a review see \citet{lo}.

The luminous extragalactic masers, with {\wlum}\,$>$\,10\,{\solum}, are classified as ``megamasers'' and are mostly found in galaxies with Seyfert~2 or LINER characteristics, exhibiting very high X-ray absorbing column densities ($N_{\rm H}$\,$>$\,10$^{23}$\,cm$^{-2}$; \citealt{zhang}). So far, Very Long Baseline Interferometry (VLBI) studies have been reported for about 10 megamaser sources and all indicate that the maser emission is located within a few parsecs from the associated active galactic nucleus (AGN), being related to (1) nuclear accretion disks (e.\,g., NGC\,4258, \citealt{n4258gree}, \citealt{n4258miyo}; IC\,2560, \citealt{ic2560}; Mrk\,1419, \citealt{mrk1419}), (2) interactions between the jet(s) emerging from the  nucleus and ambient molecular clouds (e.\,g., NGC\,1052, \citealt{n1052}; NGC\,1068, \citealt{galli}; Mrk\,348, \citealt{mrk348}), and (3) nuclear outflows (Circinus, \citealt{circinus}).

The less luminous extragalactic masers, the so-called ``kilomasers'', with {\wlum}\,$<$\,10\,{\solum}, are usually found in star-forming galaxies and the majority of them are associated with star-forming regions clearly offset from the nucleus (e.\,g., M\,33, \citealt{m33}; M\,82, \citealt{m82}; IC\,10, \citealt{hen86}; IC\,342, \citealt{ic342}; NGC\,2146, \citealt{n2146}). However, in 3 galaxies, M\,51 \citep{m51}, NGC\,4051 \citep{n4051}, and NGC\,253 \citep{n253}, kilomaser emission is also arising from the nuclear region, with projected distances from the putative nucleus of $<$\,15, 5, and 6\,pc, respectively. Among these, only the kilomaser in M\,51 has been definitely associated with an AGN, most likely with the radio jet \citep{hagiwara07}. Instead, the kilomaser in NGC\,253 turned out to be associated with a supernova remnant in the proximity of the nucleus \citep{n253new}. The nature of the water maser emission in NGC\,4051 is  still uncertain and needs to be investigated at higher resolution \citep{hagiwara07}. The possibility that some kilomasers could be related to nuclear activity (like M\,51 and, possibly NGC\,4051), similarly to the more powerful megamasers, is exciting and would provide a new tool for studying their host galaxies' nuclear regions. Hence, given the small number (15) of kilomasers found so far, this motivates extended searches to find more of these sources.
\paragraph{The sample:} Searching for 22\,GHz water vapor masers toward a sample comprised of all galaxies with declination $\delta > -30${\degr} and 100\,$\mu$m IRAS point source flux densities S$_{\rm 100\,\mu m}$ $>$\,50\,Jy, led to four new detections \citep[ hereafter HPT]{n2146, ic342, newmas}, yielding a total detection rate of 22\%. A relationship between FIR flux density and likelihood of detectable water maser emission was also established. Motivated by these results, we have decided to extend the observational campaign toward lower infrared fluxes, to confirm the correlation found and to discover additional maser sources suitable for interferometric follow-up observations.
The sample includes all galaxies with $\delta > -30${\degr} and IRAS point source flux densities 30\,Jy$<$\,S$_{\rm 100 \mu m}$\,$<$\,50\,Jy with known recessional velocities. The list of sources, shown in Table\,\ref{survey}, was compiled using the IRAS Point Source Catalog \citep{iras}. For three of the 41 target galaxies listed, NGC\,613 \citep{n3079, braatz96}, NGC\,5135 \citep{braatz04}, and NGC\,6951 \citep{claussen2, braatz96, braatz04} upper limits had already been determined. Because of the time variability of the 22\,GHz {\water} maser emission, the sources were re-observed.
In the following, we describe the results of this new survey.


\section{Observations and data reduction}\label{obs}
\subsection{Effelsberg observations}

\begin{table*}
\centering
\caption{List of galaxies searched for {\water} maser emission.}
\label{survey}
\begin{minipage}[t]{\textwidth}
\centering
\renewcommand{\footnoterule}{}
\begin{tabular}{lccrcrccc}

\hline\hline

Source                  & RA\footnote{Coordinates and velocities are taken from the NASA/IPAC Extragalactic Database (NED). The systemic velocities are given by $V_{\rm sys} = cz$. The exception is NGC\,613 for which the velocity of \citet{RC3} has been taken (see also Sect.~\ref{sect:ngc613}).}           & Dec$^a$          & $V_{\rm sys} ^a$ & $S_{\rm 100\,\mu m}$\footnote{Flux densities at 100\,${\rm \mu m}$ are taken from the IRAS Point Source Catalog \citep{iras}.} & $V$--range\footnote{Inspected velocity ranges.} & rms    & \wlum\footnote{Upper limits for maser luminosities are derived from \wlum/[\solum] = 0.023 $\times$ $S$/[Jy] $\times$ $\Delta$v/[\kms] $\times$ $D^2/[\rm Mpc^2]$, where $S$ is 3 times the rms in column 7 and $\Delta$v is the channel width, 78\,kHz or $\sim$\,1.04\,{\kms}.}  & Epoch     \\
                        &              &              &               &                     &            &        &         &           \\
                        & (J2000)      & (J2000)      & (\kms)        & (Jy)                & (\kms)     & (mJy)  & \solum  &           \\
\hline

   NGC 157              & 00 34 46.7   & -08 23 47    & 1668          &  37.73              & 1120, 2220  & 10     & $<$0.39 & 21-Jan-05  \\
   NGC 278              & 00 52 04.3   &  47 33 02    &  640          &  45.16              & 100, 1180   & 8      & $<$0.04    & 21-Jan-05  \\
\bf{NGC 520}\footnote{Source name in bold-face: newly detected maser.} & 01 24 35.1 &  03 47 33 & 2281 & 48.40 & see Table \ref{lines}&  & & \\
\bf{NGC 613}$^e$      & 01 34 18.2   & -29 25 07    & 1475          &  49.14              & see Table \ref{lines} &            &                  & \\
   NGC 908              & 02 23 04.6   & -21 14 02    & 1509          &  44.43              & 960, 2060   & 13     & $<$0.40    & 21-Jan-05  \\
   NGC 1385             & 03 37 28.3   & -24 30 05    & 1493          &  35.62              & 950, 2040   & 22     & $<$0.65    & 19-Jan-05  \\
   UGC 2866             & 03 50 14.9   &  70 05 41    & 1232          &  49.99              & 690, 1780   & 14     & $<$0.29    & 06-Apr-04  \\
   NGC 1482             & 03 54 38.9   & -20 30 09    & 1916          &  46.52              & 1370, 2470  & 19     & $<$0.96    & 10-Apr-05  \\
   NGC 1614             & 04 33 59.8   & -08 34 44    & 4778          &  32.38              & 4210, 5340  & 14     & $<$3.44    & 12-Apr-04  \\
   NGC 2339             & 07 08 20.5   &  18 46 49    & 2206          &  30.24              & 1660, 2760  & 9      & $<$0.59    & 19-Jan-05 \\
   NGC 2566             & 08 18 45.7   & -25 30 03    & 1637          &  39.43              & 1130, 2220  & 30     & $<$1.12    & 12-Apr-04  \\
   NGC 3175             & 10 14 42.1   & -28 52 19    & 1101          &  31.44              & 560, 1650   & 27     & $<$0.44    & 21-Jan-05  \\
   NGC 3310             & 10 38 45.9   &  53 30 12    &  993          &  41.76              & 450, 1540   & 13     & $<$0.17    & 06-Apr-04 \\
   MESSIER 95           & 10 43 57.7   &  11 42 13    &  778          &  35.30              & 230, 1320   & 9      & $<$0.07    & 21-Jan-05 \\
   NGC 3504             & 11 03 11.2   &  27 58 21    & 1534          &  32.89              & 990, 2080   & 8      & $<$0.24    & 21-Jan-05 \\
   NGC 3593             & 11 14 37.0   &  12 49 05    &  628          &  35.79              & 90, 1170    & 8      & $<$0.04    & 21-Jan-05  \\
   NGC 3675             & 11 26 08.6   &  43 35 09    &  770          &  35.20              & 230, 1310   & 6      & $<$0.05    & 21-Jan-05  \\
   NGC 3810             & 11 40 58.8   &  11 28 17    &  993          &  31.40              & 450, 1540   & 9      & $<$0.12    & 31-Mar-05 \\
   NGC 3893             & 11 48 38.2   &  48 42 39    &  967          &  35.07              & 420, 1510   & 9      & $<$0.11     & 31-Mar-05 \\
   NGC 4030             & 12 00 23.6   & -01 06 00    & 1465          &  46.34              & 920, 2010   & 10     & $<$0.28    & 21-Jan-05 \\
   NGC 4041             & 12 02 12.2   &  62 08 14    & 1234          &  31.49              & 690, 1780   & 14     & $<$0.29    & 06-Apr-04  \\
   NGC 4157             & 12 11 04.4   &  50 29 05    &  774          &  43.24              & 230, 1320   & 8      & $<$0.07    & 10-Apr-05  \\
   NGC 4355             & 12 26 54.6   & -00 52 39    & 2179          &  33.78              & 1630, 2730  & 12    & $<$0.79    & 19-Jan-05 \\
   NGC 4536             & 12 34 27.1   &  02 11 16    & 1808          &  44.98              & 1260, 2360  & 8     & $<$0.36     & 31-Mar-05  \\
   NGC 4605             & 12 39 59.4   &  61 36 33    &  143          &  30.53              & -400, 680   & 13    & $<$0.003     & 06-Apr-04  \\
   NGC 4654             & 12 43 56.6   &  13 07 35    & 1037          &  35.26              & 490, 1580    & 11    & $<$0.16     & 10-Apr-05  \\
   UGC 8058             & 12 56 14.2   &  56 52 25    &12642          &  30.89              & 12030, 13250 & 11   & $<$21.57    & 06-Apr-04  \\
   NGC 5033             & 13 13 27.5   &  36 35 38    &  875          &  44.53              & 330, 1420   & 11    & $<$0.11    & 19-Jan-05 \\
   NGC 5078             & 13 19 50.0   & -27 24 36    & 2168          &  32.76              & 1620, 2720  & 30    & $<$1.93     & 12-Apr-04  \\
   NGC 5135             & 13 25 44.0   & -29 50 01    & 4112          &  30.83              & 3550, 4670  & 30    & $<$6.22     & 11-Apr-04 \\
   NGC 5248             & 13 37 32.1   &  08 53 06    &  1153         &  43.97              & 610, 1700   & 10    & $<$0.19     & 21-Jan-05 \\
   MESSIER 101          & 14 03 12.6   &  54 20 57    &   241         &  30.72              & -300, 780   & 17    & $<$0.01     & 06-Apr-04 \\
   NGC 5676             & 14 32 46.8   &  49 27 28    &  2114         &  30.61              & 1560, 2670  & 9     & $<$0.57     & 21-Jan-05  \\
   NGC 5713             & 14 40 11.5   & -00 17 21    &  1899         &  37.22              & 1350, 2450  & 19    & $<$0.92     & 21-Jan-05  \\
   NGC 5775             & 14 53 57.6   &  03 32 40    &  1681         &  45.21              & 1130, 2230  & 7     & $<$0.26     & 31-Mar-05  \\
   NGC 5907             & 15 15 53.7   &  56 19 44    &   667         &  36.01              & 120, 1210   & 11    & $<$0.07     & 07-Apr-04  \\
   NGC 6643             & 18 19 46.4   &  74 34 06    &  1484         &  31.78              & 940, 2050   & 15    & $<$0.44     & 10-Apr-05  \\
   NGC 6951             & 20 37 14.1   &  66 06 20    &  1424         &  37.14              & 880, 1970   & 12    & $<$0.33     & 10-Apr-05  \\
   NGC 7469             & 23 03 15.6   &  08 52 26    &  4892         &  35.22              & 4330, 5460  & 8     & $<$2.34     & 12-Apr-04 \\
   NGC 7541             & 23 14 43.9   &  04 32 04    &  2678         &  39.93              & 2120, 3230  & 11    & $<$1.05     & 12-Apr-04  \\
   NGC 7771             & 23 51 24.9   &  20 06 43    &  4277         &  38.78              & 3710, 4840  & 11    & $<$2.47     & 12-Apr-04  \\

\hline

\end{tabular}
\end{minipage}
\end{table*}

Measurements in the $6_{16} - 5_{23}$ transition of ortho-{\water} (rest frequency 22.23508\,GHz) were carried out
with the the MPIfR 100-m telescope at Effelsberg\footnote{The Effelsberg 100-m telescope is a facility of the MPIfR 
(Max-Planck-Institut f{\"u}r Radioastronomie) in Bonn.}, Germany, between April 2004 and November 2005.
The full width to half power (FWHP) beamwidth was $\approx40{\arcsec}$ and the pointing accuracy was, on average, 4{\arcsec}
and always better than 20{\arcsec}. Throughout the survey, we used a K-band high electron mobility transistor (HEMT) receiver
with two orthogonally linearly polarized channels and a rotating feed that was switched between two fixed position separated
by $\sim$\,2{\arcmin} at a frequency of 1\,Hz.
The spectrometer was an 8192 channel correlator (AK 90) in which the signal was split into 8 bands (IFs) of width 40\,MHz and 512 channels
each, yielding a channel spacing of $\sim$\,78\,kHz ($\sim$\,1.04\,\kms). 
Four intermediate frequency bands (``IFs'') were centered at a frequency corresponding to the local standard of rest
(LSR, optical convention) recessional velocity  of each galaxy, while the others had frequency offsets of $\pm$\,20\,MHz.
On January 21, 2005 we also took a high spectral resolution spectrum of NGC\,520
using 2 bands of 20\,MHz and 4096 channels each, yielding a spacing of $\sim$\,5\,kHz (0.07\,\kms).

To establish a flux density scale, the continuum sources NGC\,7027, 3C\,286, and W3(OH) were observed 
(for flux densities, see \citealt{n7027} and \citealt{w3oh}).
All data were reduced using standard procedures with the CLASS program of the GILDAS software package.
For each source, all scans at the same frequency were added to reduce the noise in the final spectrum.
All spectra were calibrated to convert the intensity scale from noise diode units to flux density (in Jy),
using an updated version of Eq.\ 1 of \citet{galli}, which also takes into account gain variations as a function of elevation:
\begin{eqnarray*}\label{eqn:cal}
\frac{S_{\nu}}{[{\rm Jy}]} & = & \frac{N}{\rm [counts]} \times \frac{2.5}{[{\rm Jy\,counts^{-1}}]} \times \\
& \times &  \left \{ \begin{array}{ll}
\left[ 1 - 0.01(30^{\circ} - \theta^{\circ})\right]^{-1},  & \textrm{if $\theta^{\circ} < 30^{\circ}$,}\\ 
1.0,                                                                        &  \textrm{if $30^{\circ} < \theta^{\circ} < 60^{\circ}$,}\\ 
\left[1 - 0.01(\theta^{\circ} - 60^{\circ})\right]^{-1},  &  \textrm{if $\theta^{\circ} > 60^{\circ}$,}
\end{array} \right.
\end{eqnarray*}
where:
\begin{itemize}
\item N is the signal measured by the Effelsberg telescope
\item the third term corrects for the elevation dependent gain of the telescope ($\theta^{\circ}$: elevation in degrees).
\end{itemize}
The flux calibration uncertainty is estimated to be of the order of $\pm$10\%.
More observational details are given in Table \ref{survey}.

\subsection{VLA observations of NGC\,520, archival data, and imaging}\label{sect:vlaobs}

\begin{table*}
\centering
\caption{Interferometric continuum maps of NGC\,520.}
\label{maps}
\begin{minipage}[t]{\textwidth}
\centering
\renewcommand{\footnoterule}{}
\begin{tabular}{lcccccccc}

\hline\hline

Band    & $\nu$ & Observation  & Telescope & Weighting & Restoring Beam  & Position        & $\theta_{\rm max}$\footnote{Largest angular scales that can be imaged.} & rms \\
        & (GHz) &   date       &           &           & HPBW (\arcsec)  & Angle ($\degr$) & (\arcsec)          & (mJy/$\Omega_b$) \\
\hline
L       & 1.4    & 10-Dec-2000 & MERLIN    &  Natural                &$0.350\times 0.350$   & 0.00   & 3  &  0.1  \\
C       & 5.0    & 19-Nov-2004 & MERLIN    &  Tapered 1.3 M$\lambda$ &$0.187\times0.123$    & --51.12 & 1  &  0.07 \\
X       & 8.5    & 27-Dec-2004 & VLA-A     &  Uniform                &$0.223\times0.215$    & 18.89  & 7  &  0.03 \\
U       & 14.9   & 10-May-1986 & VLA-A     &  Natural                &$0.205\times0.151$    & 19.81  & 4  &  0.1  \\
K       & 22.1   & 14-Feb-2005 & VLA-BnA   &  Natural                &$0.369\times0.136$    & 63.21  & 7  &  0.5  \\
K       & 22.5   & 17-Apr-2006 & VLA-A     &  Natural                &$0.112\times0.084$    & 11.29  & 1  &  0.06  \\
\hline
\end{tabular}
\end{minipage}
\end{table*}

NGC\,520 was observed at 22\,GHz with the NRAO\footnote{ The National Radio Astronomy Observatory is a facility of the National Science Foundation operated under cooperative agreement by Associated Universities, Inc.} Very Large Array (VLA) in the BnA and A configuration on February 14, 2005 and April 17, 2006, respectively. In February 2005 observations were made in Spectral Line Mode, using a single bandwidth of 3.1\,MHz, which corresponds to a velocity coverage of $\sim$\,45\,{\kms}, centered at V$_{\rm LSR}$\,=\,2276\,{\kms}.
The use of 256 spectral channels yielded a spacing of 12.2\,kHz, corresponding to 0.16\,{\kms}.
In April 2006, NGC\,520 was observed both in Spectral and Continuum Mode.
For the spectral line observations, we employed the set up used in February 2005, while continuum observations were made using
two bands of 50\,MHz each, centered at frequencies of 22.4851 and 22.4351\,GHz.
``Referenced Pointing'' observations were made in regular intervals to warrant good antenna
pointing. The ``Fast Switching'' technique was used with a typical duty cycle of 4 minutes, $\sim$170\,sec on the target source
(NGC\,520) and $\sim$70\,sec on the phase calibrator 0121+043 (0.67\,Jy).
The flux density scale was established by observing the non-variable source 0137+331 (3C48; 1.13\,Jy). \\
Archival VLA A-array continuum data at 8.5\,GHz (project AA296, P.I. Hofner) and 14.9\,GHz (project AC156, P.I. Carrol) were also reduced by us. These observations were taken in December 2004 and May 1986, respectively. Absolute calibration was established at 8.5\,GHz using 3C48 (3.15\, Jy) and at 14.9\,GHz using 3C286 (3.45\, Jy).
The compact sources 0125-000 (1.15\, Jy) and 0106+013 (3.11\, Jy) were measured to calibrate the telescope phases.
Furthermore, we have produced two continuum MERLIN\footnote{MERLIN is operated by the University of Manchester as a National facility of the Science and Technology Facilities Council (STFC).} maps at 1.4 and 5.0\,GHz.
NGC\,520 was observed at 1.4\,GHz and 5.0 GHz with the full MERLIN array in December 2000 and November 2004, respectively. For the 1.4 GHz session, the 76-m Lovell telescope took also part in the observations.

All datasets were calibrated in the standard way, except that, since 3C\,48 is resolved by the VLA both at 8 and 22\,GHz, phase and amplitude correction factors were determined by using CLEAN component models for this source. Each individual dataset was Fourier-transformed using natural weighting (with the exception of the 8.5\,GHz map where we used uniform weighting) and then deconvolved using the CLEAN algorithm \citep{clean}. Table~\ref{maps} illustrates the details of the continuum maps.

%

\section{Results}\label{res}

Toward two of the 41 observed targets, the Seyfert galaxy NGC\,613 and the merging system NGC\,520, {\water} maser emission was detected. Figs. \ref{fig:ngc613} and \ref{fig:ngc520} show the line profiles, while line parameters are given in Table \ref{lines}. Below, we summarize the main properties of the two galaxies and their masers.

\begin{table*}
\centering
\caption{Line parameters of newly detected masers.}
\label{lines}
\begin{minipage}[t]{\textwidth}
\centering
\renewcommand{\footnoterule}{}
\begin{tabular}{lccccrrrccc}
\hline\hline

Source  & Adopted  & Epoch  & Tel. & $\int{S{\rm d}V}$\footnote{Integrated flux densities, center velocities, and the full width to half maximum (FWHM) linewidths of the observed features are obtained from Gaussian fits.}  & $V_{\rm LSR,opt}^a$ & $\Delta V_{1/2}^a$ & $V$-range\footnote{Inspected velocity ranges.} & Channel & rms   & \wlum\footnote{Isotropic luminosities are derived from \wlum/[\solum] = 0.023 $\times$ $\int{S {\rm d}V}$/[Jy\,\kms] $\times$ $D^2/[\rm Mpc^2]$.}  \\
             & Distance\footnote{Distances have been estimated using $H_0$=75\,{\kms}\,Mpc$^{-1}$.} &              &     &                     &                                    &                          &                      &  Width      &           &            \\
             & (Mpc)  &       &     &(mJy\,{\kms}) & ({\kms})                    & ({\kms})           & ({\kms})     & ({\kms})  & (mJy) & ({\solum})    \\

\hline

NGC 613 & 20       & Jan 19 2005 &  Eff &1500$\pm$300   & 1453$\pm$2        &  23$\pm$4     & 930, 2020   & 4.3     & 18    & 14      \\
        &          &             &      &  2300$\pm$300 & 1502$\pm$3        &  37$\pm$6     &             & 4.3     & 18    & 21      \\
NGC 520 & 30       & Jan 21 2005 &  Eff & 120$\pm$10    & 2275.28$\pm$0.05  & 1.1$\pm$0.6   &  1730, 2830 & 1.1     & 8     &  2.1    \\
        &          &             &      &  70$\pm$60    & 2275.42$\pm$0.08  & 0.6$\pm$0.2   &  2010, 2560 & 0.3     & 28    &  1.4    \\
        &          & Feb 14 2005 & VLA  &  25$\pm$2     & 2275.15$\pm$0.03  & 0.61$\pm$0.07 &  2250, 2300 & 0.2     & 4     & 0.5     \\
        &          & Mar 31 2005 & Eff  &   \dots       &     \dots         &    \dots      &  2010, 2560 & 0.3     & 29    & $<$0.5    \\
        &          & Apr 10 2005 & Eff  &   \dots       &     \dots         &    \dots      &  1730, 2830 & 1.1     & 11    & $<$0.8   \\
        &          & Nov 14  2005 & Eff &   \dots       &     \dots         &   \dots       &  2010, 2560 & 1.1     & 8      & $<$0.5 \\
        &          & Apr 17 2006 & VLA  & \dots         &   \dots           &  \dots        &   2250, 2300 & 0.2     & 5      & $<$0.1 \\
\hline
\end{tabular}
\end{minipage}
\end{table*}

\subsection{NGC\,613}\label{sect:ngc613}

NGC\,613 is a barred spiral galaxy \citep{RC3}, located at a distance of $\sim$\,20\,Mpc (1\,{\arcsec} $\cor$ 100\,pc, for $H_0$=75\,{kms}\,Mpc$^{-1}$). Optical spectroscopy revealed that this galaxy has a composite spectrum, with an \HII\, region and a Seyfert-like contribution simultaneously present \citep{veron}. High resolution radio continuum maps of the inner kpcs of NGC\,613 show a linear structure, consisting of three distinct blobs, oriented perpendicular to a ring-like feature of radius $\sim$\,350\,pc \citep{hummel}. A large amount of CO has been detected \citep{n613co} but high angular resolution images of molecular emission have, so far, not yet been reported.

\begin{figure}
   \resizebox{\hsize}{!}{\includegraphics[angle=-90]{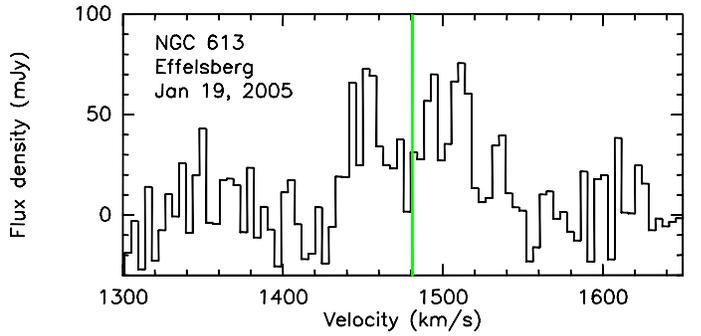}}
      \caption{\water\, megamaser profile observed toward NGC~613 with Effelsberg on January 19, 2005. Channel spacing and pointing position are 4.3~{\kms} and $\alpha_{2000}$ = 01$^{\rm h}$ 34$^{\rm m}$ 18$^{\rm s}$.2, $\delta_{2000}$ = --29$^{\circ}$ 25\arcmin\ 07\arcsec, respectively. The velocity scale is with respect to the local standard of rest (LSR) and uses the optical convention. The vertical line indicates the {\HI} systemic velocity $V_{\rm sys}=1481\pm5$\,{\kms} (NED).}
         \label{fig:ngc613}
   \end{figure}

The \water\, maser profile in NGC\,613, shows two distinct components (see Fig. \ref{fig:ngc613}). The most luminous one is centered at $\sim$\,1500\,{\kms}, which is redshifted w.r.t. the systemic velocity of the galaxy, and has a full width to half maximum linewidth of $\sim$\,40\,{\kms}. A second blueshifted component is seen at $\sim$\,1450\,{\kms} and has a linewidth of $\sim$\,20\,{\kms}. The total isotropic maser luminosity is $\sim$ 35\,{\solum}. Independently from our observation, the maser was detected by \citet{kon} in July 2003. They report a position for the maser coincident with the optical position of the nucleus with an uncertainty of 1\,{\farcs}3, supporting the association of the maser emission with AGN activity. The comparison of our spectrum, from 2005, with that of \citet{kon} (their Fig.~1), from 2003, shows that the line profiles are similar. However, in the former, the peak flux density of the component at $V_{\rm LSR} \sim$\,1450\,{\kms} appears to be higher. The large linewidth and the presence of not more than two main components, separated by only a few 10\,{\kms}, may hint at a jet-maser, as proposed by \citet{kon}. Nevertheless, the {\water} emission is centered on the {\HI} systemic velocity (see Fig.~\ref{fig:ngc613}) and on the CO centroid velocity (1465\,{\kms}; \citealt{n613co}), which is not common for jet-masers. Characteristically, jet-masers are entirely found at one side of the systemic velocity, either blue-  or redshifted \citepalias{newmas}. 

\subsection{NGC\,520}\label{sect:ngc520}

NGC\,520 is a nearby (D$\sim$\,30\,Mpc, 1\arcsec\ $\cor$ 145\,pc, for $H_0$=75\,{kms}\,Mpc$^{-1}$) merging system, considered to be the result
of an encounter between a gas rich and a gas poor disk galaxy \citep{n520HvG}.
The remnant nuclei are clearly visible in near infrared K-band (2\,${\rm \mu m}$) images
\citep{n520stan90,n520koti01}. At optical wavelengths, the primary nucleus, which is four times more massive than the other
\citep{n520stan90}, is hidden behind a pronounced dust lane, while the secondary less massive one 
is seen 40{\arcsec} (6\,kpc) to the north-west \citep{n520stan90,n520laine03}.

The radio continuum emission arising from NGC\,520 is confined to a linear east-west structure of $\sim$\,6\arcsec ($\sim$\,870\,pc) in extent, with the peak coinciding with the 2\,${\rm \mu m}$ position of the main nucleus \citep{condon82,n520carral90,n520besw03}. The 4.9\,GHz unresolved peak source of \citet{condon82} is resolved into 3 components in the 14.9\,GHz map of \citet{n520carral90}. The 14.9\,GHz map reveals a total of 5 components, aligned  along the linear structure. \citet{n520besw03} imaged NGC\,520 at 1.4 and 1.6\,GHz with MERLIN. In their highest resolution map, the linear continuum structure resolves into 10--15 compact components. Five of these coincide with the sources detected by \citet{n520carral90}, which \citet{n520besw03} associated with bright supernova remnants (SNRs) and radio supernovae (RSNe).

The inner 1\,kpc surrounding the primary nucleus contains a large concentration of neutral atomic and molecular gas. {\HI} and OH have been detected in absorption, along with an OH maser feature \citep{n520besw03}. The velocity distribution of both {\HI} and OH trace a kpc-scale ring or disc at a position angle (P.A.) of 90{\degr} rotating about the main nucleus with a velocity gradient of 0.5\,km\,s$^{-1}$\,pc$^{-1}$. This result is consistent with the CO velocity field \citep{n520yun01}.

{\it Chandra} X-ray observations \citep{n520xray} show no significant diffuse emission from the secondary nucleus, while
diffuse emission, consistent with a starburst driven wind, has been discovered outflowing from the primary nucleus,
perpendicular to the molecular CO disk \citep{n520yun01}.
{\it Chandra} observations also revealed 15 point sources within the optical boundaries of the galaxy \citep{n520xray}.
The third brightest of these, source 12, is an ultraluminous X-ray source (ULX) with
$L_{\rm X}$(0.2--10\,keV)\,$\sim 9 \times$10$^{39}$\,erg\,s$^{-1}$. Since it coincides, within the uncertainty,
with the radio peak \citep{condon82}, the center of the molecular disk traced by CO \citep{n520yun01},
and the center of the diffuse wind structure, it has been associated with the main nucleus of NGC\,520.

\subsubsection{The detection and variability of the {\water} maser}\label{sect:n520maser}

\begin{figure}
   \resizebox{\hsize}{!}{\includegraphics{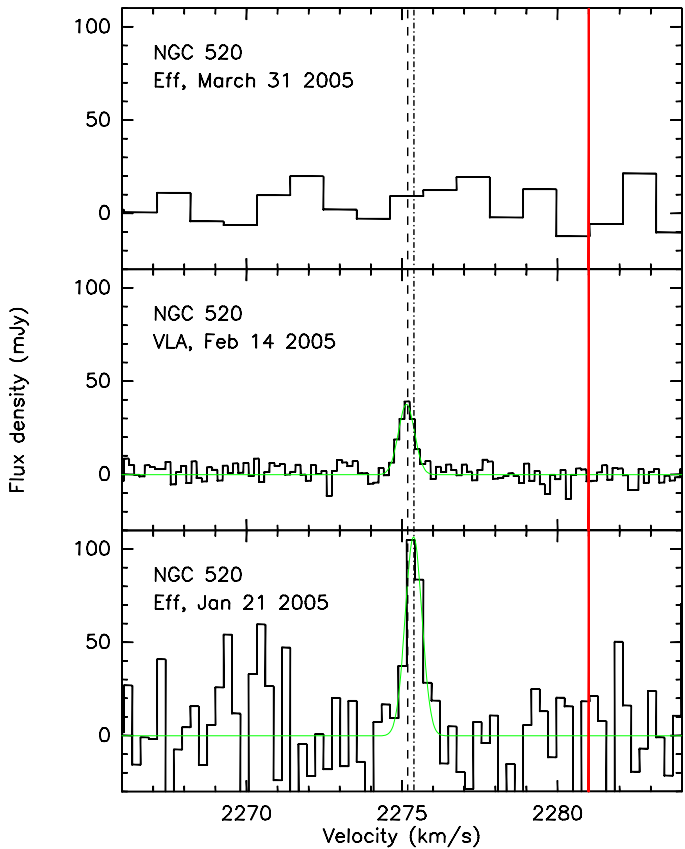}}
      \caption{\water\, kilomaser spectra observed toward NGC~520 between January 21 and March 31, 2005. Channel spacings are 0.6~{\kms} (lower panel), 0.2~{\kms} (middle panel), and 1.1~{\kms} (upper panel). The pointing position is $\alpha_{2000}$ = 01$^{\rm h}$ 24$^{\rm m}$ 35$^{\rm s}$.1, $\delta_{2000}$ = +03$^{\circ}$ 47\arcmin\ 33\arcsec. The solid vertical line indicates the systemic velocity $V_{\rm sys}$=2281\,{\kms} (NED). The dash-dotted and dashed lines represent the peak velocities of the maser in the Effelsberg and VLA spectra, respectively.}
         \label{fig:ngc520}
   \end{figure}

On January 21, 2005, we detected a single, narrow {\water} maser line in NGC\,520 (see Fig~\ref{fig:ngc520}). The line parameters are reported in Table~\ref{lines}. The inferred isotropic maser luminosity is $\sim$\,1\,{\solum}. On February 14, the VLA observations confirmed the presence of a single, narrow emission feature. As is evident from Fig.~\ref{fig:ngc520} and Table~\ref{lines}, the peak flux density of the line has decreased w.r.t. the earlier Effelsberg spectrum. In addition, a small shift of $\sim$ 0.3 {\kms} toward the blue is apparent. A 2-dimensional Gaussian fit (made using AIPS task JMFIT) to an image of the channel maps showing maser emission indicates a spatially unresolved spot at a position $\alpha_{2000}$=01$^{\rm h}$24$^{\rm m}$34\fs914$\pm$0\fs001 and $\delta_{2000}$=03\degr47\arcmin29\,{\farcs}79$\pm$0\,{\farcs}01.
The uncertainties of the position are the formal errors determined by JMFIT. These are consistent with the theoretical error given, for each coordinate, by $\sim$\,0.5$\theta_b$/$SNR$, where $\theta_b$ is the beam (HPBW) size along either RA or Dec and SNR is the signal-to-noise ratio in the map.  Having used the `Fast Switching' technique (see Sect.~\ref{sect:vlaobs}), the absolute positional accuracy is likely consistent with this value. Nevertheless, for the  absolute uncertainty we conservatively take  0\,{\farcs}1, the canonical error for VLA K-band A-array observations.
The position and the velocity of the {\water} maser are incompatible with those of the weak OH masers detected by \citet{n520baan85} and \citet{n520besw03}.

In the Effelsberg observation made on March 31, no emission above the 3$\sigma$ noise level of $\sim$\,90\,mJy was found in the 0.3\,{\kms} wide channels (Fig.~\ref{fig:ngc520} shows a smoothed spectrum). The fading of the maser line was confirmed by our most recent spectrum, taken with the VLA on April 17, 2006, with a 3$\sigma$ detection threshold of 15\,mJy for a 0.2\,{\kms} wide channel.
 The decrease in flux density of $\sim$\,70\% within 24 days (from January 21 to February 14), implies a size scale $\la$\,6.2$\times$10$^{16}$\,cm ($\la$\,0.02\,pc) or $\la$ 0.14\,mas at a distance of 30 Mpc (if variations are intrinsic to the masing cloud(s)). The corresponding brightness temperature is $\ga 10^{10}$\,K.

We produced continuum maps from the VLA line data cubes of February 2005 and April 2006 using the line free channels and find an unresolved spot with a peak flux density of 3--4\,mJy\,beam$^{-1}$ at the position of the maser. We compared these maps with the broad band continuum map at 22.1\, GHz (see next Sect.). Position and flux density of this source are not compatible with those of the continuum source detected in the broad band map. In order to investigate this ambiguity, we smoothed the line data cubes to a resolution of 0.8\,{\kms} to increase the signal-to-noise ratio in individual channels. In these maps we see ``tentative'' weak line emission at a 3--4$\sigma$ level at exactly the same position as the main maser line but at different velocities. We made additional continuum maps excluding these velocity channels and the  continuum source is still present at the same position, though with lower flux density. Hence, we think that the spot seen in the line-free maps is likely caused by the blending of weak maser lines below the detection threshold of the individual channel maps, possibly spread over the entire 45\,{\kms} bandpass. 

\subsubsection{The 22\,GHz continuum source}\label{sect:csou}

\begin{figure}
\resizebox{\hsize}{!}{\includegraphics{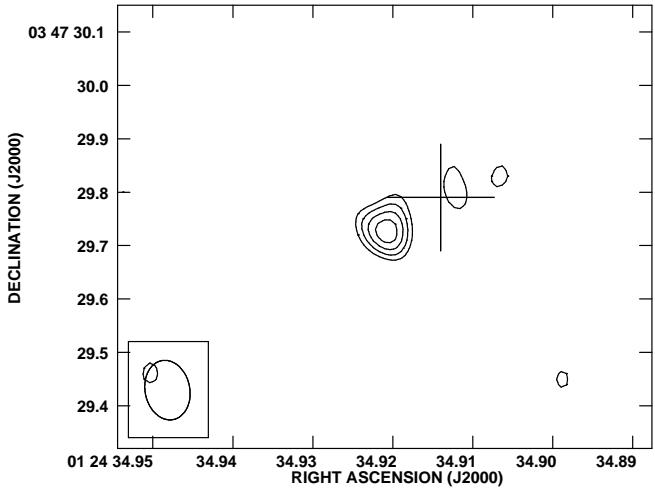}}
\caption{VLA A-array 22.5\,GHz broad band continuum map of NGC\,520, from April 17, 2006.
               Contour levels are -3, 3, 4, 5, 6 $\times$ 0.06\,mJy\,beam$^{-1}$.
                The cross marks the position of the {\water} maser.}
\label{ksource}
\end{figure}

In the 22.5\,GHz continuum image made from the broad band continuum dataset, we find a single, almost unresolved source (Fig.~\ref{ksource}, hereafter labeled ``K''), whose position, determined  fitting an elliptical Gaussian with JMFIT, is $\alpha_{2000}$=01$^{\rm h}$24$^{\rm m}$34\fs921$\pm$0\fs001 and $\delta_{2000}$=03\degr47\arcmin29\,{\farcs}73$\pm$0\,{\farcs}01. If we assume, as before, that the absolute positional accuracy is 0\,{\farcs}1, then the uncertainty on the relative positions is 0\,{\farcs}14. Within this conservatively estimated error the maser and the continuum source coincide. However, as we stated in Section~\ref{sect:n520maser}, the absolute position error is probably smaller then this value and might even be of order 10\,mas. In this case the peaks of the maser and that of source K do not coincide any longer (they are displaced by 0\,{\farcs}12$\pm$0\,{\farcs}014, or 18$\pm$2\,pc).
The peak and integrated flux densities of source K are 0.41$\pm$0.07\,mJy\,beam$^{-1}$ and 0.5$\pm$0.1\,mJy, respectively.
The undeconvolved extent is 110$\pm$20$\times$100$\pm$20\,mas with position angle (P.A.) $\sim$\,+30\degr, matching the size and P.A. of the beam. The brightness temperature of the source is $\ga$\,40\,K. 

\subsubsection{Nuclear continuum sources}\label{sect:nsou}

\begin{figure*}
\centering

\includegraphics[scale=0.7]{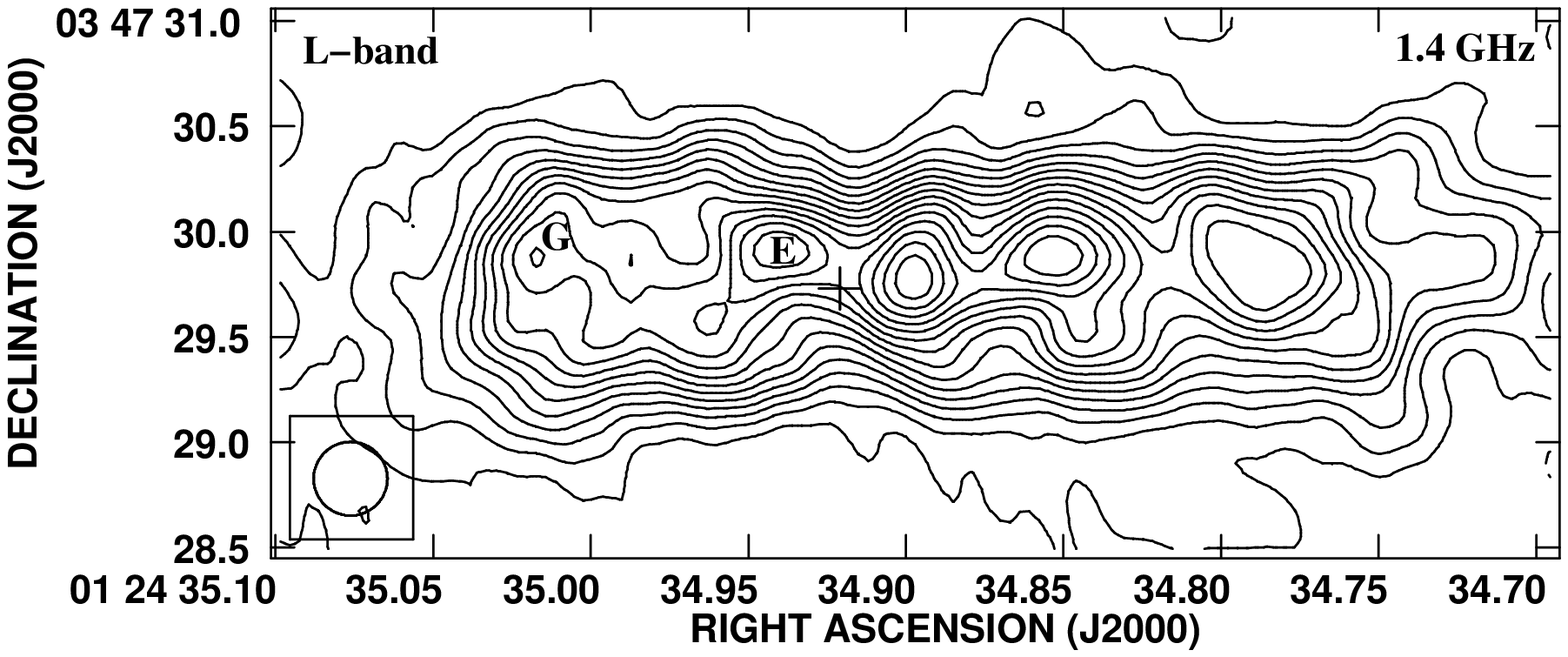}

\includegraphics[scale=0.7]{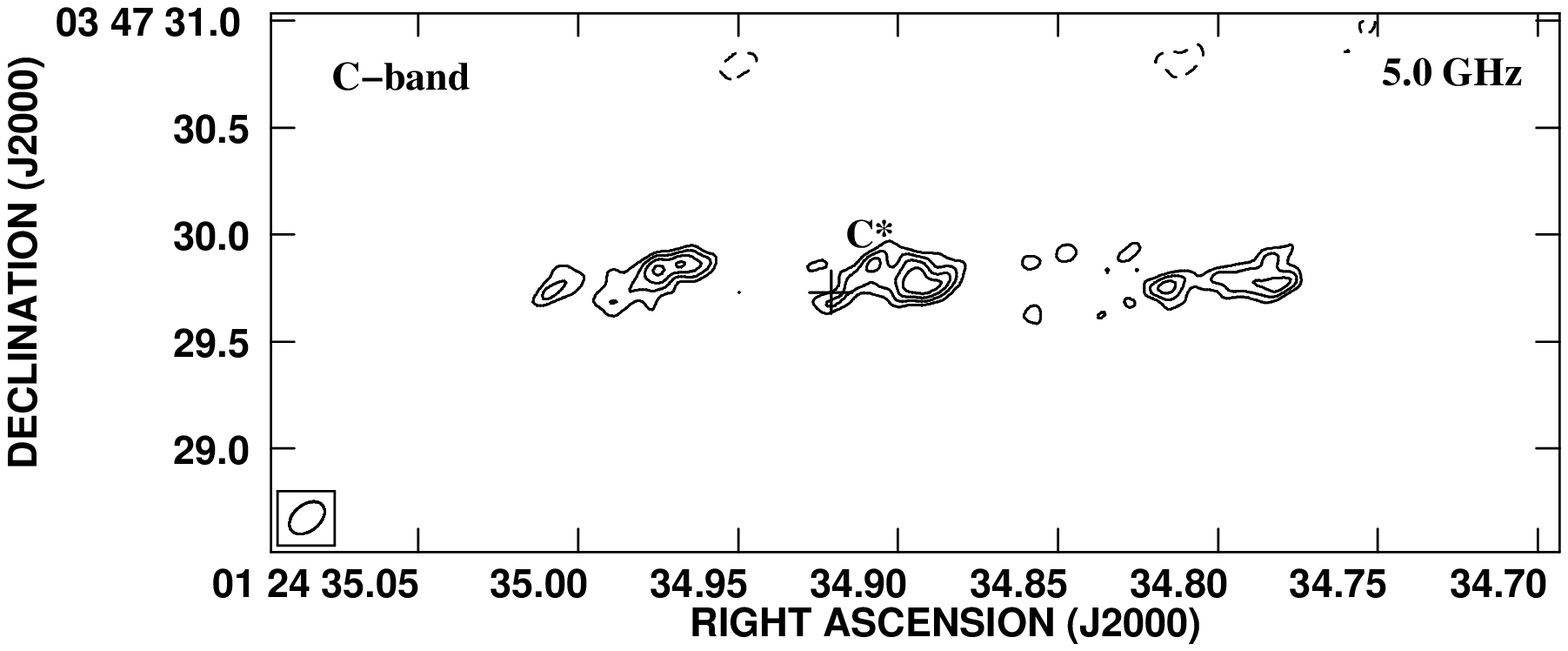}

\includegraphics[scale=0.7]{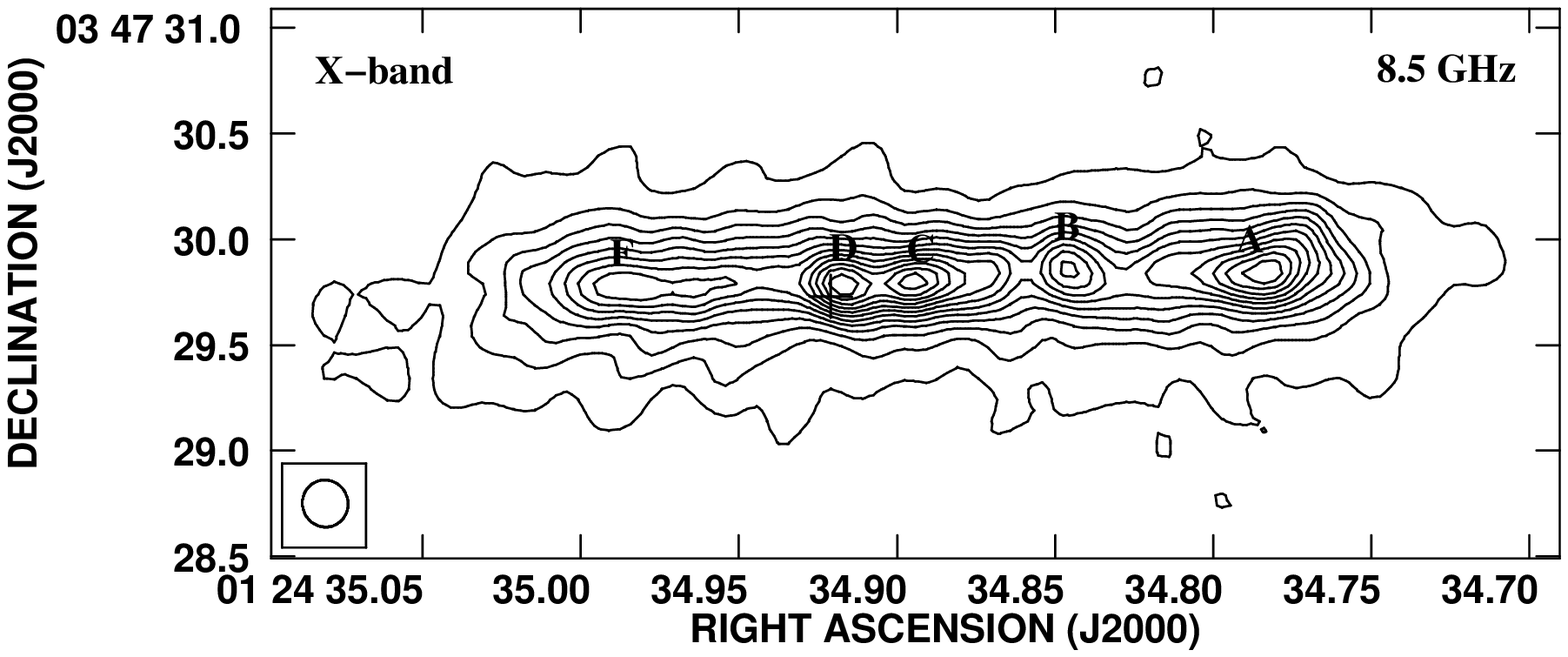}

\includegraphics[scale=0.7]{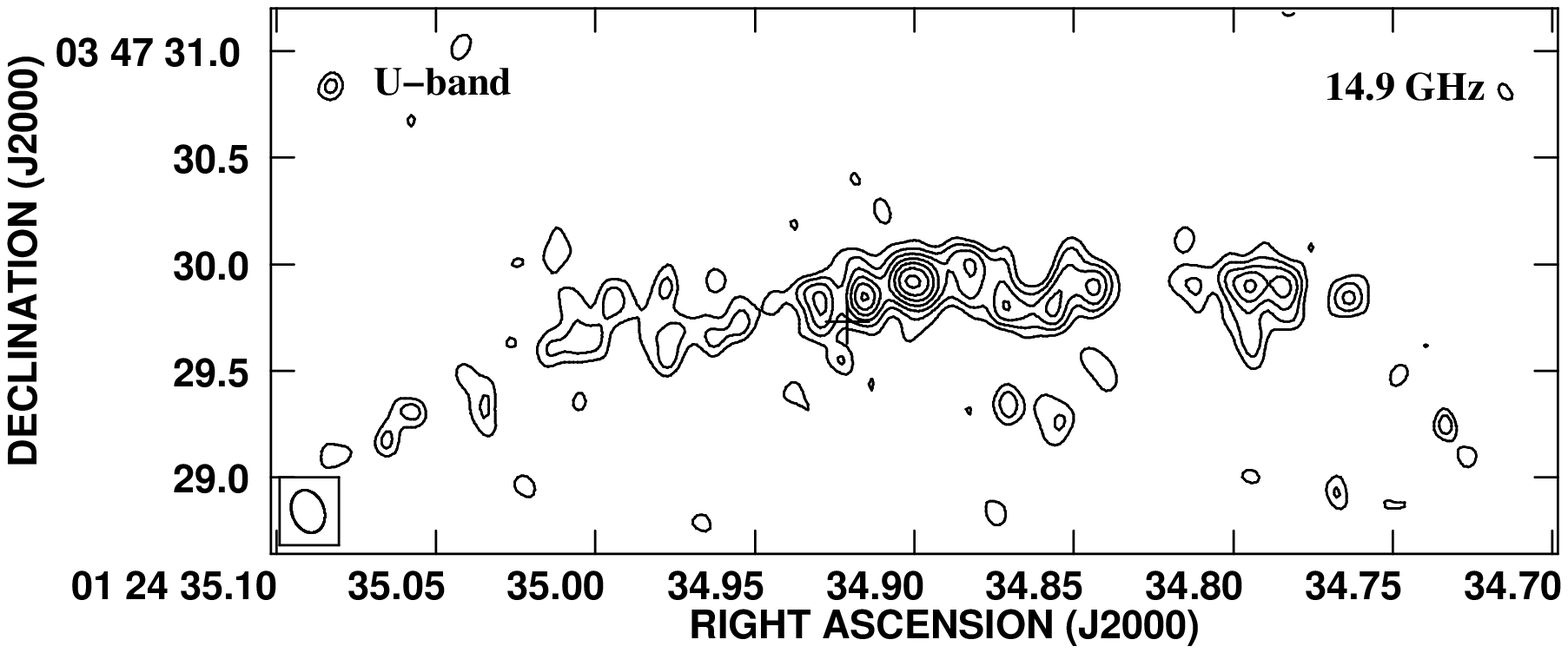}

\caption{Radio continuum contour maps of the nuclear region of NGC\,520 with labels indicating the compact sources A-G. From top to bottom: L-band, MERLIN, contour levels are --3, 3, 6, 9, 12, 15, 18, 21, 24, 27, 30, 33, 36, 39, 42, 45, 48 $\times$\,0.1\,mJy\,beam$^{-1}$; C-band, MERLIN, contour levels are --3, 3, 4, 5, 6 $\times$ 0.1\,mJy\,beam$^{-1}$; X-band, VLA, contour levels are --1, 1, 2, 3, 4, 5, 6, 7, 8, 9, 10, 11, 12, 13, 14, 15, 16 $\times$ 0.1\,mJy\,beam$^{-1}$; U-band, VLA, contour levels are --3 3 4 5 6 7 8 9 $\times$ 0.1\,mJy\,beam$^{-1}$. The cross marks the position of continuum source K detected at 22\,GHz (Fig.~\ref{ksource}).}
\label{kntr}
\end{figure*}

Figure \ref{kntr} shows the MERLIN and VLA images of the inner 6\arcsec$\times$2\,{\farcs}5 region surrounding
the main nucleus of NGC\,520 at the frequencies of 1.4, 5.0, 8.5, and 14.9\,GHz (details of these maps are given in Table ~\ref{maps}).
Consistent with previous studies (e.g. \citealt{n520carral90} and \citealt{n520besw03}), a linear east-west structure is visible at all frequencies. This structure is resolved into several compact components, the number of which depends on the resolution and sensitivity of the maps. 
In our maps it is possible to identify a total of 7 compact sources, labeled with the letters A to G.
The four sources A, B, C, and F, are resolved into sub-components in the MERLIN C-band and VLA U-band maps, because of their higher resolution. Coordinates, peak and integrated flux densities are displayed in Table~\ref{merlin_fluxes} for the MERLIN maps and in Table~\ref{vla_fluxes} for the VLA data. From a comparison with previously published radio continuum maps, we can associate our five sources A -- E with those labeled C1 -- C5 of \citet{n520carral90} and 4 -- 8 of \citet{n520besw03}. It is, however, not possible to associate the other two sources, F and G, with any of the sources reported by \citet{n520carral90} or \citet{n520besw03}.


\begin{table*}
\begin{minipage}[t]{\textwidth}
\caption{Properties of the nuclear continuum sources in NGC~520 observed with MERLIN at 1.4 and 5 GHz.}
\label{merlin_fluxes}
\centering
\begin{tabular}{lcccccc}
\hline\hline

Label & $\alpha$ & $\delta$ & \multicolumn{2}{c}{$\nu = 1.4$GHz} &  \multicolumn{2}{c}{$\nu = 5.0$GHz} \\
          & (J2000)   & (J2000)  & Peak flux  & Total flux & Peak flux & Total flux \\
          & 01$^h$24$^m$ & 03\degr47\arcmin & (mJy/$\Omega_{\rm b}$) & (mJy) & (mJy/$\Omega_{\rm b}$) & (mJy) \\

\hline

A & 34\fs78 & 29\,{\farcs}8 & 3.5$\pm$0.1  & 4.7$\pm$0.3 & 0.56$\pm$0.08 & 1.0$\pm$0.2   \\
  & 34\fs82 & 29\,{\farcs}8 &              &             & 0.55$\pm$0.07 & 0.55$\pm$0.07 \\
B & 34\fs85 & 29\,{\farcs}9 & 4.5$\pm$0.1  & 4.5$\pm$0.1 & 0.38$\pm$0.07 & 0.38$\pm$0.07 \\
  & 34\fs86 & 29\,{\farcs}9 &              &             & 0.35$\pm$0.07 & 0.35$\pm$0.07 \\
C & 34\fs90 & 29\,{\farcs}8 & 4.8$\pm$0.1  & 5.4$\pm$0.4 & 0.69$\pm$0.08 & 2.3$\pm$0.3   \\
C$^*$  & 34\fs91 & 29\,{\farcs}9 &              &             & 0.54$\pm$0.07 & 0.54$\pm$0.07 \\
D & 34\fs92 & 29\,{\farcs}8 & $<$0.3       & \dots          & $<$0.2        & \dots            \\
E & 34\fs93 & 29\,{\farcs}8 &  4.2$\pm$0.1 & 4.2$\pm$0.1 & $<$0.2        &     \dots        \\
F & 34\fs97 & 29\,{\farcs}9 & 3.4$\pm$0.1  & 3.4$\pm$0.1 & 0.62$\pm$0.07 & 0.62$\pm$0.07 \\
  & 34\fs98 & 29\,{\farcs}8 &              &             & 0.62$\pm$0.07 & 0.62$\pm$0.07 \\
G & 35\fs01 & 29\,{\farcs}7 & 3.3$\pm$0.1  & 3.3$\pm$0.1 & 0.44$\pm$0.07 & 0.44$\pm$0.07 \\

\hline

\end{tabular}
\vfill
\end{minipage}
\end{table*}

\begin{table*}
\begin{minipage}[t]{\textwidth}
\caption{Properties of the nuclear continuum sources in NGC~520 observed with the VLA at 8 and 15 GHz.}
\label{vla_fluxes}
\centering
\begin{tabular}{lcccccccc}
\hline\hline

Label & $\alpha$ & $\delta$ & \multicolumn{2}{c}{$\nu = 8.5$GHz} & \multicolumn{2}{c}{$\nu = 14.9$GHz} & $\alpha^8_{15, \rm p}$ &  $\alpha^8_{15, \rm i}$ \\
          & (J2000)   & (J2000)  & Peak flux & Total flux & Peak flux & Total flux &  Spectral index & Spectral index \\
          & 01$^h$24$^m$ & 03\degr47\arcmin & (mJy/$\Omega_{\rm b}$) & (mJy) & (mJy/$\Omega_{\rm b}$) & (mJy) & (from peaks) & (from integrated fluxes)\\

\hline

A & 34\fs78 & 29\,{\farcs}8 & 1.18$\pm$0.03  & 5.7$\pm$0.2   & 0.7$\pm$0.1  & 2.2$\pm$0.1 & -0.8$\pm$0.2 & -1.5$\pm$0.4 \\
B & 34\fs85 & 29\,{\farcs}9 & 0.97$\pm$0.03  & 0.97$\pm$0.03 & 0.6$\pm$0.1  & 0.6$\pm$0.1  &  -0.8$\pm$0.3 & -0.8$\pm$0.3 \\
  & 34\fs86 & 29\,{\farcs}9 &                &               & 0.6$\pm$0.1  & 0.6$\pm$0.1 \\
C & 34\fs90 & 29\,{\farcs}8 & 1.27$\pm$0.03  & 4.2$\pm$0.1   & 0.9$\pm$0.1  & 2.5$\pm$0.4  &  -0.5$\pm$0.2 & -0.8$\pm$0.3 \\
D & 34\fs92 & 29\,{\farcs}8 & 1.29$\pm$0.03  & 1.29$\pm$0.1   & 0.8$\pm$0.1  & 0.8$\pm$0.1  &  -0.8$\pm$0.3 & -0.8$\pm$0.3 \\
E & 34\fs93 & 29\,{\farcs}8 &  $<$0.1        &  \dots           & 0.6$\pm$0.1  & 0.6$\pm$0.1  &  & \\
F & 34\fs97 & 29\,{\farcs}9 & 0.90$\pm$0.03  & 5.5$\pm$0.2   &   $<$0.3     & \dots           &  & \\
G & 35\fs01 & 29\,{\farcs}7 &  $<$0.1        & \dots            &  $<$0.3      & \dots           &  & \\

\hline

\end{tabular}
\vfill
\end{minipage}
\end{table*}

For all maps, with the exception of the MERLIN L-Band one, the parameters (peak coordinates and peak and integrated flux densities; Tables~\ref{merlin_fluxes} and ~\ref{vla_fluxes}) were derived fitting a 2-dimensional Gaussian (AIPS task: JMFIT). In the L-Band map, positions were measured using the AIPS task IMSTAT, which reads a portion of an image and provides the location and value of the maximum brightness in that portion. Integrated flux densities were obtained using the task BLSUM, which sums the pixels' flux densities over blotch polygonal regions. The blotch areas over which to integrate where determined according to the average dimensions of the source (as derived from the other maps) in order to avoid confusion with the extended background present in the L-Band map. The flux density uncertainties reported in the two tables (\ref{merlin_fluxes} \& \ref{vla_fluxes}) are provided directly by the tasks used and range between a few percent up to 20 percent of the values measured.

At 6\,cm, the longest baseline, $B_{\rm max}$, of MERLIN (215 km) provides a resolution of $\sim$ 0.05{\arcsec} that is a factor of 6 better than that of the VLA in its most extended configuration. Therefore, MERLIN maps can resolve a source seen by the VLA as one in two or more components (e.g. our sources A and C). In addition, since the shortest baseline, $B_{\rm min}$, of MERLIN is 9 km, while that of the VLA A-array is 0.68 km, the two instruments have quite a different response to extended structure (see also Table~\ref{maps}): at 6 cm, the MERLIN interferometer is ``blind'' to emission more extended than 1{\arcsec} while the VLA can map emission on a 10{\arcsec}--15{\arcsec} scale. This latter problem, which is obvious when comparing our 5.0\,GHz MERLIN map with the others (Fig.~\ref{kntr}), will affect the computation of the peak flux densities and integrated fluxes. The background of the MERLIN L-Band and the VLA maps is resolved-out in the C-Band where, rather, a bowl-like depression is created, which reduces the values of the peak flux densities \citep[see e.\,g., ][]{wills98}. Furthermore, some extended emission in the larger of the compact sources may also be missed in the calculation of the integrated fluxes. For this reason, we have decided not to use the 5.0 GHz MERLIN data in our computation of the spectral indices. In addition, due to the lower resolution of the L-Band MERLIN images some sources are blended, and hence, we have also not used this image for the spectral index estimates. Nevertheless, it is useful to show the MERLIN maps in order to compare the global morphology of the radio emission from NGC~520 at different frequencies and to potentially spot new sources (e.g. new radio supernovae) that might serendipitously appear in galaxies with high star formation rates.

Spectral indices were then computed from the VLA maps at 8.5 and 14.9\,GHz following the convention $S \propto \nu^{\alpha}$. The resolution of the two maps is almost identical, and therefore, a convolution to the same beam was not necessary.
The last two columns of Table~\ref{vla_fluxes} show the computed spectral indices using both peaks and integrated fluxes. The errors on the spectral index $\alpha$ were derived using the standard equation for error propagation,
\begin{displaymath}
\Delta \alpha = \pm \frac{0.434}{\lg \nu_2 - \lg \nu_1} \sqrt{\Big(\frac{\Delta S_{\nu_1}}{S_{\nu_1}}\Big)^2 + \Big(\frac{\Delta S_{\nu_2}}{S_{\nu_2}}\Big)^2},
\end{displaymath}
where $\nu$, $S_{\nu}$, and $\Delta S_{\nu}$ are the frequency, flux density, and error of the flux density, respectively. The subscripts `1' and `2' refer to frequencies 8.5 and 14.9\,GHz, correspondingly.

%

\section{Discussion}\label{disc}

\subsection{The FIR sample}

The FIR sample of intermediate bright galaxies (30$<$\,S$_{\rm 100\,\mu m}$\,$<$\,50\,Jy)
comprises 40 spirals and a galaxy pair (NGC\,520). Among them, 14 are starburst galaxies and only 10 are classified as Seyferts.
As for the FIR sample of brighter galaxies \citepalias[$S_{\rm 100\,\mu m} >$\,50\,Jy; see ][]{newmas}, there are very few galaxies with
FIR luminosities greater than 10$^{11}$\,{\solum} in our sample. There are only four Luminous Infrared Galaxies (LIRGs, with
$L_{\rm FIR}>$10$^{11}$\,{\solum}), NGC\,1614, NGC\,5135, NGC\,7469, and NGC\,7771 and one Ultraluminous Infrared Galaxy
(ULIRG, $L_{\rm FIR}>$10$^{12}$\,{\solum}), UGC\,8058.

\begin{figure}
   \resizebox{\hsize}{!}{\includegraphics{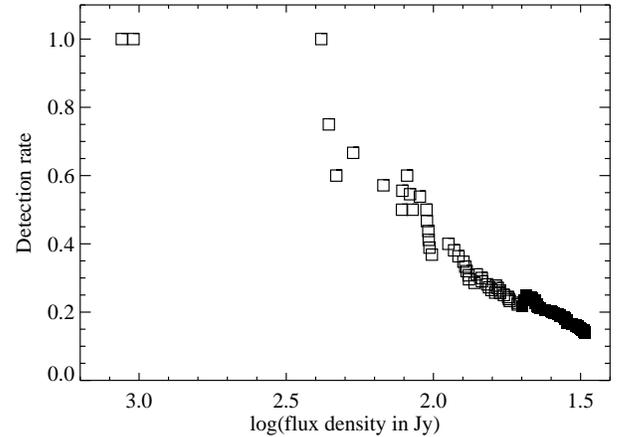}}
      \caption{{\water} maser cumulative detection rate above a certain 100 $\rm {\mu m}$ IRAS Point Source Catalog flux density for the parent galaxy. The figure is a modified version of Fig.~13 of \citetalias{newmas}. Filled squares represent our values derived using the ``intermediate-luminosity'' sample.}
         \label{ratemod}
\end{figure}

Adding the result of our `intermediate-range' sample, i.e. two new maser detections (NGC~613 and NGC~520) out of 41 galaxies searched, to that of \citetalias{newmas}, the cumulative detection rate above a 100 $\rm {\mu m}$ IRAS Point Source Catalog flux of 30 and 40\,Jy for the parent galaxy becomes 14\% and 20\%, respectively (Fig.~\ref{ratemod}). When considering the detection rate related to the specific FIR flux density interval (30--50 Jy) of our sample, we obtain 2/41 or 5\%. The two detected galaxies, NGC\,613 and NGC\,520, are the second and third FIR-brightest galaxies of the sample. The FIR-brightest galaxy of the sample, UGC\,2866, thus might also be detectable if frequent monitoring is performed (\textit{nota bene} the high degree of variability we see in NGC 520!). 

\begin{figure}
   \resizebox{\hsize}{!}{\includegraphics{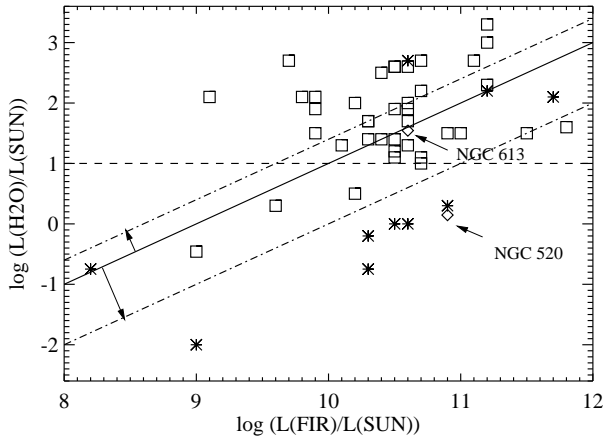}}
      \caption{{\water} vs. FIR luminosity plot including our two new maser detections, NGC\,613 and NGC\,520, and the separation between the kilomaser and megamaser regimes (dashed line). Open squares represent most of the 22\,GHz extragalactic {\water} masers detected so far \citepalias{newmas}. The asterisks indicate the remaining 10 maser sources belonging to the FIR subsample with $S_{\rm 100\,\mu m} >$\,50\,Jy \citepalias[see ][]{newmas}. The solid line marks the correlation for Galactic star-forming regions \citep{jaffe}. The upper and lower dot-dashed diagonal lines show the FIR/{\water} luminosity relations for the mega- and kilomasers, respectively, with the slope being fixed to that of the \citet{jaffe} correlation.}
         \label{kilomega}
\end{figure}

\citetalias{newmas} found a rough proportionality between  $L_{\rm FIR}$ and {\wlum} for extragalactic water masers. When plotting the value for NGC~613 into this diagram (Fig.~\ref{kilomega}), it follows the relationship representing the $L_{\rm FIR}$/{\wlum} correlation for Galactic masers found by \citet{jaffe}. The point representing NGC\,520 is, however, located well below the line of the correlation. This system has either an excess in $L_{\rm FIR}$ w.r.t. {\wlum} or its maser luminosity is particularly low. Most likely, the bulk of the 22\,GHz {\water} emission remains undetected because of limitations in sensitivity (see the Sect,~\ref{sect:n520maser}). When taking a closer look at the $L_{\rm FIR}$/{\wlum} plot and considering megamaser and kilomaser sources separately (above and below the dashed {\wlum} = 10\,{\solum} dividing line), the two groups seem to follow independent correlations. While {\water} kilomasers are ``subluminous'' for a given $L_{\rm FIR}$ w.r.t. Galactic star-forming regions \citep{jaffe}, the megamasers appear to be slightly ``superluminous''. The two samples diverge by $\sim$\,1.5 orders of magnitude in maser luminosity for a given FIR luminosity, emphasizing their different origin.

\subsection{NGC\,520}

\subsubsection{Where is the nucleus?}\label{sourcek}

\begin{figure}
   \resizebox{\hsize}{!}{\includegraphics{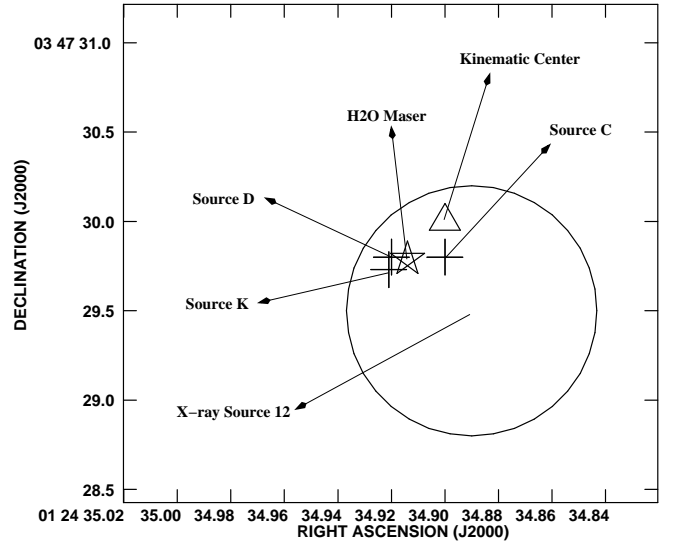}}
      \caption{Positions of the main centers of activity in the nuclear region of NGC~520. The plus signs mark the radio continuum sources C, D, and K detected in our VLA maps, while the five-pointed star indicates the position of the \water\ maser (this paper). The triangle identifies the kinematic center derived from CO and \HI\ studies \citep{n520yun01}. The circle indicates the luminous X-ray source (labeled 12) detected by \citet{n520xray} using Chandra. The sizes of the marks denote absolute positional accuracies.}
         \label{nucleo}
\end{figure}

Our measurements place the kilomaser in NGC\,520 in the innermost nuclear region of the galaxy. In order to associate the maser emission with either enhanced star formation or with nuclear activity, we have first to locate and identify the galaxy's true nucleus and check whether it exhibits (low-luminosity) AGN activity. So far, no clear evidence for the presence of an AGN in NGC\,520 has been reported. It is impossible to determine the spectral type of the main nucleus on the basis of its optical spectrum because of a lack of diagnostic emission lines \citep{veilleux95}. This is likely due to the large extinction suffered by the main nucleus (see Sect.~\ref{sect:ngc520}). However, an indication of absent (or weak) nuclear activity is the presence of PAH features in the mid-infrared spectrum \citep{brandl06}, which should be diluted by the strong infrared continuum of an AGN \citep{krolik01}. No Fe-K line seems to be present in the X-ray spectrum but an ULX source (source 12, see Sect.~\ref{sect:ngc520}) with a ``power law spectrum'' ($N_{\rm H}$=2.3 $\times$10$^{22}$\,cm$^{-2}$) is detected close to the main nucleus \citep{n520xray}. Fig.~\ref{nucleo} shows a plot in which the positions for the main centers of activity are marked. The X-ray source 12 lies closest to radio source C (see Sect.~\ref{sect:nsou}) and, according to \citet{n520xray}, is ``undoubtedly associated with the primary nucleus of NGC~520''. However, due to the uncertainty of Chandra's absolute astrometry the error circle ($\approx$ 0\,{\farcs}7) also encompasses sources D and K. Which is, then, the true nucleus of the galaxy?

Source K is close to sources C (further resolved into a double source in the MERLIN C-Band image) and D. Assuming the conservative accuracy of 0\,{\farcs}1 (\,$\equiv$\,15\,pc at a distance of 30 Mpc; see Fig.\ref{kntr}) for the absolute positions in all the VLA maps, source K cannot be associated with source C, but an association with D is possible (the angular distance is $\sim$\,0\,{\farcs}07; see Fig.~\ref{nucleo}). The slightly-resolved source C has a spectral index between 8.5 and 14.9\,GHz, $\alpha^8_{15}$, of $-0.5 \pm 0.2$ (from the peak flux densities) and of $-0.8 \pm 0.3$ (from the integrated fluxes; see Table \ref{vla_fluxes}). If the slope of the spectrum still holds at higher frequencies, it is possible to derive the expected integrated intensities for source C at 22.5\,GHz, $1.8 \pm 0.4$\,mJy. In order to make the integrated fluxes for source C and K consistent, a spectral index $\alpha^{15}_{22}$ of $\approx - 4$ would be necessary, which is impossible to justify with standard physical models. For source D, the spectral index is $\alpha^8_{15} = -0.8 \pm 0.3$, providing an expected value for the 22.5\,GHz integrated intensity of $0.6 \pm 0.1$\,mJy. This value agrees within the uncertainties with that of source K (0.5$\pm$0.1\,mJy; see Sect.~\ref{sect:csou}), making the association between the two sources, D and K, the most convincing option. Given this interpretation, it remains quite surprising that no other source, and especially source C, is detected at 22.5\,GHz. The non-detection of source C must be a consequence of its angular extent, confirmed also by the MERLIN C-Band map. The other sources are likely too weak to be detected in our highest frequency map due to their non-thermal nature.

\begin{figure}
   \resizebox{\hsize}{!}{\includegraphics{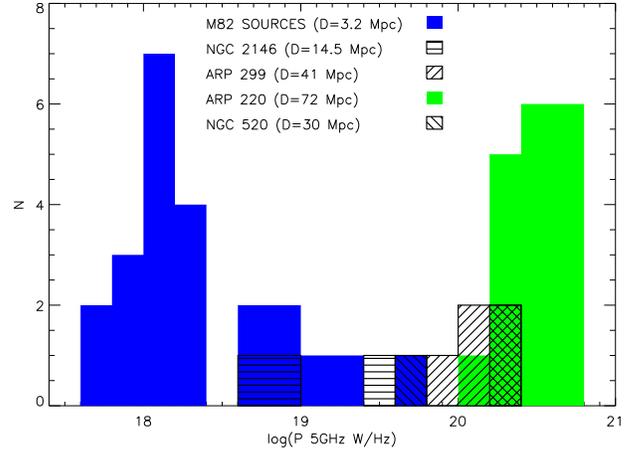}}
      \caption{5\,GHz luminosity distribution of the non thermal sources C, C$^*$, and D (or K) (for this source the flux density has been extrapolated from the 8 GHz VLA map using the spectral index reported in Table.~\ref{vla_fluxes}) in NGC\,520 (this paper), compared with the SNRs in M\,82 \citep{muxlow94}, NGC~2146 (Tarchi 2001, Ph.D. Thesis), Arp\,299 \citep{neff04}, and Arp\,220 \citep{parra07}.}
         \label{isto}
\end{figure}

\begin{figure}
   \resizebox{\hsize}{!}{\includegraphics{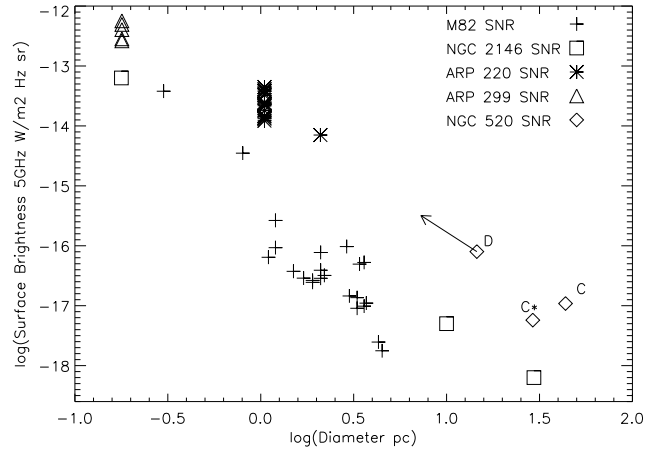}}
      \caption{$\Sigma$-D relation for SNRs in M\,82 (crosses; \citealt{muxlow94}), NGC\,2146 (squares; Tarchi 2001, Ph.D. Thesis and Surcis, 2004, Ph.D. Thesis), Arp\,299 (triangles; \citealt{neff04}), Arp\,220 (asterisks; \citealt{parra07}), and the sources C, C$^*$, and D (or K) (diamonds) in NGC~520 (this paper). The arrow shows the direction into which source D would be moved, if its size would be $<$15\,pc.}
         \label{sigmad}
\end{figure}

What, then, is the nature of source K (or D)? 
Its size is smaller than 0\,{\farcs}1 ($\equiv$ 15 pc) and its spectrum is characterized by a steep radio spectral index, which is consistent with optically thin synchrotron emission from a RSN or a young SNR, similar to the situation found in other starburst galaxies like M~82 \citep[e.\,g., ][]{mcdonald02} and NGC~2146 \citep{tarchi00}. The luminosity of source K (or D) at 5 GHz (extrapolating the flux density from the 8.5\,GHz value using a spectral index of --0.8; Table~\ref{vla_fluxes}) is $\approx$ 2$\times$10$^{20}$\,W\,Hz$^{-1}$, which is greater than the luminosity of the strongest non-thermal compact source found in M~82, but comparable with the luminosities of SNRs found in other merger systems, like Arp\,299 and Arp\,220 (Fig.~\ref{isto}). However, when plotting the surface brightness of source K ($\sim$\,8 $\times$ 10$^{-17}$ W\,Hz$^{-1}$\,m$^{-2}$\,sr$^{-1}$)  as a function of its diameter (15 pc) in the $\Sigma{\rm -D}$ diagram for the SNRs in M\,82, NGC\,2146, Arp\,299 and Arp\,220 (Fig.~\ref{sigmad}), the value is incompatible with those obtained for the other sources (with the exception of the SNRs in Arp\,220). Being spatially unresolved, the derived source size is an upper limit, and hence, the position of source K might move toward the upper-left corner of the plot, where the strongest sources in M~82, NGC~2146, and Arp\,299 lie. Nevertheless, the anomalous position of source K in the $\Sigma$-D plot might also suggest that it is a low luminosity AGN (LLAGN). Indeed, radio observations of Seyfert and LINER galaxies show that LLAGN may have steep radio spectral indices \citep{morganti99, krips07} and luminosities comparable to that of source K \citep{morganti99}. In summary, the parameters derived for  source K (or D), viz.\, spectral indices, luminosities and diameter, are consistent both with those of SNRs and LLAGNs. Multi-epoch and/or multi-frequency VLBI observations would be necessary to select between the two options. 

Source C has also a steep spectral index, a large luminosity (2.3 $\times$ 10$^{20}$\,W\,Hz$^{-1}$ at 5.0\,GHz) and a position in the $\Sigma$-D diagram which does not match those of the other sources (Fig.~\ref{sigmad}). However, one of the key points that determine its position in the aforementioned diagram is its size, which we estimate to be of order 0\,{\farcs}3 ($\equiv$40 pc). This size is too large to justify a possible LLAGN nature, and hence, we think that source C is a particularly  active star-forming region composed of a collection of small and young SNRs/RSNe that we do not fully resolve. This is also suggested by our C-Band MERLIN map that shows the presence of at least one further emission spot (C$^*$, Table~\ref{merlin_fluxes}) close to what we identify as source C. This spot has a size $\la$0\,{\farcs}2 ($\equiv$ 30\,pc), a 5\,GHz luminosity of $\approx$ 5.5 $\times$ 10$^{19}$\,W\,Hz$^{-1}$, and a surface brightness $\ga$ 5.7 $\times$ 10$^{-18}$\,W\,m$^2$\,Hz$^{-1}$\,sr$^{-1}$. These properties are more consistent with those of SNRs (see Figs.~\ref{isto} and \ref{sigmad}), supporting our interpretation. Also, the kinematic center, although close-by, is not coincident with source C favoring a different view from that suggested by \citet{n520xray}. Given this, it still remains unclear where the nucleus of the galaxy is located, but if a LLAGN exists in NGC\,520, it is likely associated with source D.

\subsubsection{The {\water} maser in NGC~520: star formation or nuclear activity?}

The association of the maser with source K (or D) is based on the assumption that the uncertainty in the absolute positions is 0\,{\farcs}1 in our 22\, GHz VLA maps (see Sect.~\ref{sect:csou}). If a less conservative value is assumed, line and continuum do not coincide anymore. Therefore, the maser might  not be directly associated with any of the continuum sources. Nevertheless, the small projected distance between the maser spot and source K suggests that the {\water} emission is in some way related to this source. 
The characteristics of the maser line, i.e. narrowness and strong time variability (70\% decrease of the flux density within 24 days) are more consistent with star formation than with nuclear activity. Narrow flaring components are common in water maser emission from Galactic star-forming regions, although only the outbursts in W49N reached a luminosity comparable to that of NGC\,520 \citep{w49n_lij,w49n_honma}. Narrow extragalactic {\water} maser lines have been detected, so far, also in IC\,10 \citep{hen86} and IC\,342 \citep{ic342}. Both these kilomasers are definitely associated with star-forming regions and have shown a great deal of time variability. The flare of the $-324${\kms} component in IC\,10 reached $\sim$\,4\,{\solum} or 120\,Jy \citep{ic10baan} only to decay to sub-Jy levels more recently \citep{ic10brunthaler}. Even more similar to the {\water} maser in NGC~520 is the maser in IC~342 reported by \citet{ic342}. Although less luminous (\wlum\ = 10$^{-2}$ \solum), the feature was narrow ($\sim$\,0.6\,{\kms}) and varied by a factor of 3 within 5 days \citep{ic342}. Moreover, for the maser in IC~342, a velocity shift by 1\,{\kms} was observed within two weeks. The shift and the rapid decay in flux was explained in terms of a chance alignment of two dense molecular clouds. 

A similar scenario is also compatible with the maser in NGC\,520. When comparing our highest velocity resolution spectra taken with Effelsberg on January 2005 with that taken with the VLA a month later, a small velocity shift seems present. A Gaussian fit to the maser lines indicates velocities of 2275.42 $\pm$ 0.08 and 2275.15 $\pm$ 0.03 for the Effelsberg and the VLA, respectively (Fig.~\ref{fig:ngc520}). The velocity shift is 0.27$\pm$0.09\,{\kms}, i.\,e. it may be significant. We may assume that both the flare and the shift (if real) have a kinematic origin, following the scenario proposed by \citet{ic342}. Assuming a negligible distance between the two clouds w.r.t. the distance of the galaxy and a relative velocity in the plane of the sky of $\sim$\,100\,{\kms}, this implies a cloud velocity gradient of up to 0.23 \kms/AU during our monitoring time (24 days).
Alternatively, the maser feature may be explained by a change in the relative intensities of the three main hyperfine components \citep[see e.\,g., ][]{fiebig89} or by a variation in the relative intensity of different emitting regions leading to a swapping-over of velocity peaks \citep[see e.\,g., ][]{wu99}. For maser flux density variations (distinct from flares) due to refractive interstellar scintillations, time scales range from about one month (slightly longer than the variability time scale in NGC\,520) to few months \citep{simonetti93}. Diffractive scintillation may instead allow for extremely rapid variations (as short as a few minutes) but has been reported, so far, in only one case, viz. the Circinus Galaxy \citep{greenhill97}.

Although the attributes of the maser (low luminosity, narrowness, and strong variability) hints at an association  with star formation, we cannot rule out the possibility that we have found a weak nuclear kilomaser, as the ones in M\,51 and (possibly) NGC\,4051, perhaps related to a nuclear jet that is too weak to be seen in our radio maps. However, no high-velocity lines blue- or redshifted with respect to the systemic velocity are detected in NGC\,520 and this does not support the possibility that the maser is tracing an accretion disk. In addition, the linewidth of the maser feature in NGC\,520 ($\sim$\,0.6\,{\kms}) is very small compared to the typical linewidth of {\water} masers associated with nuclear jets (e.\,g., Mrk\,348, \citealt{mrk348}; M\,51, \citealt{hagiwara07}). Most of all, no convincing evidence of an AGN of whatever intensity comes from any study performed so far, making the latter alternative unlikely. However, also a few megamaser galaxies, seen edge on and thus being extremely obscured (like NGC\,520), do not show the presence of an AGN \citep{n6300, braatz04}.
%

\section{Conclusions}\label{concl}

We have observed a sample of 41 galaxies with 100\,$\mu$m IRAS point source flux density between 30 and 50\,Jy and declination $>-30${\degr}, to search for 22\,GHz {\water} maser emission. The survey yielded two detections, a megamaser with isotropic luminosity, {\wlum}, of $\approx$\,35\,{\solum} in the Seyfert/{HII} galaxy NGC\,613 (see also \citealt{kon}) and a kilomaser with {\wlum}$\approx$\,1\,{\solum} in the merger system NGC\,520. Our results confirm the correlation between maser detection rate and FIR flux density discovered by \citetalias{newmas} for the sample of galaxies with $S_{\rm 100 \mu m} >$\,50\,Jy. When we compare the isotropic luminosity of extragalactic masers with the FIR luminosity of their parent galaxy, we find that kilomasers are ``subluminous'' while megamasers are ``superluminous'' for a given $L_{\rm FIR}$ w.r.t Galactic {\water} masers in massive star-forming regions. We agree with \citet{kon} that the maser in NGC\,613 is possibly associated with the nuclear jet, although the fact that the {\water} emission is centered on the systemic velocity is unusual for a jet-maser and may hint at a different origin.

The kilomaser in NGC\,520 has been observed at subarcsecond resolution with the VLA. We have found that the {\water} emission arises from the inner hundred parsecs surrounding the dominant nuclear region of the merger system. Assuming an absolute position accuracy of 0\,{\farcs}1 ($\equiv$\,15\,pc), the maser coincides with a continuum source (source K, or D). 
A spectral index analysis, as well as its general properties, i.e., its size and luminosity, indicate a young SNR or a LLAGN. The small linewidth ($\sim$\,0.6\,{\kms}) and the time variability (70\% decrease of the flux density within 24 days) of the maser feature suggest a size limit of $\la$0.02\,pc and a brightness temperature greater than $10^{10}$\,K, if the fluctuation is not caused by a variable background continuum source. The {\water} maser emission is most likely associated with star formation, like the majority of kilomasers. Narrowness and variability of the maser feature resemble those of two other kilomasers observed in IC~10 and IC~342 that are associated with star formation. We consider an association with an AGN a less viable option based on the lack of clear traces of AGN activity.

%

\begin{acknowledgements}
This work has benefited from research funding from the European Community's sixth Framework Programme under RadioNet R113CT 2003 5058187.
We wish to thank R.J. Beswick for kindly making available MERLIN 1.4 and 5.0\,GHz data on NGC\,520 and an anonymous referee for critically reading the manuscript. PC and AT wish to thank the operators at the 100-m telescope for their hospitality during the observing runs. This research has made use of the NASA/IPAC Extragalactic Database (NED), which is operated by the Jet Propulsion Laboratory, Caltech, under contract with NASA. This research has also made use of the NASA's Astrophysics Data System Abstract Service (ADS).
\end{acknowledgements}

\bibliographystyle{aa} 
\bibliography{8316} 

\end{document}